\documentstyle[12pt]{article}
\vspace{-2cm}
\begin{document}

\title{\large PHYSICAL MODEL OF SCHRODINGER ELECTRON.\\
              FAYNMAN CONVENIENT WAY IN MATHEMATICAL \\
              DESCRIPTION OF ITS QUANTUM BEHAVIOUR}

\vspace{-2.4cm}

\author{\small \bfseries Josiph Mladenov Rangelov  \\
Iistitute of Solid State Physics\,,\,Bulgarian Academy of Sciences\,,\\
72\,Tsarigradsko chaussee\,,\,1784\,,\,Sofia\,,\,Bulgaria\,.}
\date{}
\maketitle

\vspace{-1.3cm}

\begin{abstract}
The physical model (PhsMdl) of a nonrelativistic quantized Schrodinger's
electron (SchrEl) is offered. The behaviour of the SchrEl's well spread
(WllSpr) elementary electric charge (ElmElcChrg) had been understood by
means of two independent and different in a frequency and size motions. The
description of this resultant motion may be done by substitution of the
classical Wiener continuous integral with the quantized Feynmam continuous
integral. There are possibility to show by means of the existent not only
formal but substantial analogy between the quadratic differential wave
equation in partial derivatives of Schrodinger and quadratic differential
particle equation in partial derivatives of Hamilton-Jacoby that the addition
of a kinetic energy of the stochastic harmonic oscillation of some quantized
micro particles to the kinetic energy of classical motion of the same micro
particles formally determines their wave behaviour.It turns out the
stochastic motion of the quantized micro particles powerfully to break up the
smooth thin line of the classical motion of the same micro particle in many
broad cylindrically spread path. The  SchrEl participate in stochastically
roughly determined circumferences within different flats and with different
radii, with centres which are successively arranjed over short and very
disorderly orientated lines. Therefore the quantized motion of some micro
particle cannot be descripted by smooth thin well contured (focused) line,
describing the classical motion of the macro particle.

\end{abstract}

\section{\normalsize Introduction}

  A physical model (PhsMdl) \cite{JMRa},\cite{JMRb} and \cite{JMRc} of the
nonrelativistic quantized Schrodinger's electron (SchrEl) is offered in this
work. In our obvious PhsMdl the SchrEl will be regarded as some well spread
(WllSpr) elementary electric charge (ElmElcChrg), taking simultaneously part
in two independent and different in size and frequency motions: A) Some
classical motion of a Lorentz' electron (LrEl) along a smooth clear-cut thin
classical trajectory realized in a consequence of some a known interaction
(IntAct) of LrEl's over spread (OvrSpr) ElmElcChrg, magnetic dipole moment
(MgnDplMmn) or bare mass with the intensity of some external classical fields
(ClsFlds) as it is done within the Newton nonrelativistic classical mechanics
(NrlClsMch) and Maxwell-Lorentz nonrelativistic classical electrodynamics
(ClsElcDnm). B) The isotropic three-dimensional nonrelativistic quantized
((IstThrDmnNrlQnt) Furthian stochastic boson harmonic oscillations
FrthStchBsnHrmOscs) of the SchrEl as a result of the permanent electric
interaction (ElcIntAct) of its WllSpr ElmElcChrg with the electric intensity
(ElcInt) of the resultant quantized electromagnetic field (QntElcMgnFld) of
the stochastic virtual photons (StchVrtPhtns), stochasticly generated by dint
of StchVrtPhtns (\cite{JMRd}, \cite{JMRe} and \cite{JMRf}), exchanged between
the fluctuating vacuum (FlcVcm) and it. This Furthian quantized stochastic
behaviour of the SchrEl is very similar to the Brownian classical stochastic
behaviour of the ClsMicrPrt. But in a principle the exact description of the
resultant behaviour of the SchrEl owing of its participation in both the
mentioned motions could be done only by means of the nonrelativistic quantum
mechanics' (NrlQntMch) and nonrelativistic classical electrodynamics'
(ClsElcDnm) laws.

 The description of some quantized micro particle {QntMicrPrt} behaviour,
within the matrix presentation of the NrlQntMch, offered by Heisenberg,
(\cite{WH}) have been accompanied with unfounded affirmation that its unknown
motion cannot be described by dint of any its trajectory.Therefore for this
purpose one must use the matrix elements of its operator, which in a reality
presents a Fourie components of the same trajectory. Indeed, missunderstanding
the cause for incommon dualistic behaviour of QntMcrPrts lets one erroneous
applay an ansamble statistical commentary instead of a stochastic diffusible
one of the probabibly interpretation of the modul square of its orbital wave
function (OrbWvFnc) and gives some incorrect physical interpretation of the
uncertainity relations of Heisenberg (\cite{WH}). Therefore the contiuity
integral representation of the motion within the NrlQntMch have been
missinterpretated as a natural generalization of the classical space-time
trajectory. Some physical scientists have considered this representation as a
giving some possibility for the construction of some trajectoty, which is
compatible with the uncertainity relation of Heisenberg within Feynman's
continuity integral representation of the NrlQntMch. But this is very incorrect allegation as because these continuity path integrals are
calculated over all virtual possible trajectories in this area. In Feynman
mathematical formalism (\cite{FH}) the transition of some QntMicrPrt from one
space point into another one is characterized by no one trajectoty but by a
greet number of possible trajectories, each of them insert by certain own
contribution of the probability in its transition amplitude.

\section{\normalsize Physical explanation of the essence of the electron
physical model by analogy between the mathematical description of
FrthQntMicrPrt behaviour and BrmClsMicrPrt behaviour.}

In our obvious physical model (PhsMdl) \cite{JMRc} of the nonrelativistic
quantized Schrodinger's electron (SchrEl) it will be regarded as some well
spread (WllSpr) elementary electric charge (ElmElcChrg), taking
simultaneously part in two different motions : A) The classical motion of a
Lorentz' electron (LrEl) along an smooth well contoured thin classical
trajectory realized in a consequence of a some known interaction (IntAct) of
its over spread (OvrSpr) ElmElcChrg, magnetic dipole moment (MgnDplMmn) or
bare mass with the intensity of some external classical fields (ClsFlds) as
it is done in the Newton nonrelativistic classical mechanics (NrlClsMch) and
Maxwell-Lorentz nonrelativistic classical electrodynamics (ClsElcDnm). B) The
isotropic three-dimensional nonrelativistic quantized Furthian stochastic
boson harmonic oscillations (IstThrDmnNrlQnt FrthStchBsnHrmOsc) of the SchrEl
as a result of the permanent electric interaction (ElcIntAct) of its WllSpr
ElmElcChrg with the electric strength of the resultant quantized electromagnetic
field (QntElcMgnFld) of the stochastic virtual photons (StchVrtPhtns), generated
y dint of StchVrtPhtns exchanged between the fluctuating vacuum (FlcVcm) and it. This Furthian quantized stochastic
behaviour of the SchrEl is very similar to the Brownian classical stochastic
behaviour of the ClsMacrPrt.

 Indeed,it is well known that the classical motion of some Lorentz' electron
(LrEl) as a classical macro particle (ClsMacrPrt) is well described by means
of a clear-cut smooth narrow line, while the quantized motion of some
Schrodinger's electron (SchrEl) as a quantum micro particle (QntMicrPrt) is
well discribed is well described by sum of two line: the first one is a
distinct smooth thin classical line and the second one is many broad
cylinricaly spread path. The  SchrEl participates in stochastically roughly
determined circumferences oscillations within different flats and with
different radii, with centres which are successively arranjed over frequently
broken line,short and amounted by random very disorderly orientated in space
petty pieces lines. Therefore the quantized motion of some micro particle
cannot be descripted by smooth thin well contured (focused) line. Therefore
the quadratic differential wave equation of Schrodinger (QdrDfrWvEqtSchr)
(\cite{EScha}) may be obtained through an addition of the kinetic energy of
Furth quantized stochastic harmonic oscillation motion (\cite{RPF}), expressed
by the dispersion of its imaginery momentum or stochastic osmotic velosity to
the quadratic differential particle equation of Hamilton-Jacoby
(QdrDfrPrtEqtHml/Jcb). Since then a transparent survey of the behaviour of a
nonrelativistic quantized SchEl in our PhsMdls may be build by means of the
substitution of classical Wiener's continuous integral (\cite{NW}) with
quantized Feynman's continuous integral (\cite{RPF} and \cite{FH}). Therefore
it is necessary to take into consideration that Schrodinger in 1931
(\cite{ESchb}) and Furth in 1933 (\cite{RPF}) had found some formal analogy
between the quadratic differential diffusive equation of Focker-Plank
(QdrDfrDfsEqtFcrPln) :
\begin{equation} \label{a1} \frac{\partial
W}{\partial t}\,=\,div(Wv)\,-\,D\,\Delta W
\end{equation}

for the distribution function $W$ of a probability density (DstFncPrbDns) of
the free Brownian classical micro particle (BrnClsMicrPrt) in a motionless
coordinate system in a respect to one and the quadratic differential wave
equation of Schrodinger (QdrDfrWvEqtSch)
\begin{equation}
\label{a2}
\hbar \frac{\partial \Psi }{\partial t}\,=\,-\frac{\hbar ^2\Delta}{2m}\,\Psi \,+
\,V\,\Psi
\end{equation}

for an orbital wave function (OrbWvFnc) $\Psi $ of a free Furthian quantized
micro particle (FrthQntMicrPrt) in a motionless coordinate system in respect
to one.This similarity become particulary stricing at an absence of any
external forces when U = 0 and v = 0.
\begin{equation}
\label{b1}
\frac{\partial W}{\partial t}\,=\,-\,D\,\Delta W
\end{equation}

and
\begin{equation}
\label{b2}
i\,\hbar\frac{\partial\Psi }{\partial t}\,=\,-\frac{\hbar^2\Delta}{2m}\,\Psi
\end{equation}

 The unimportant distinction between two equations consists in the existence
of imaginary unit $i$ in diffusivity factor of wave equation, i.e. the if
diffusivity factor $D$ of BrnClsMcrPrt has real value, the diffusivity
factor $D$ has imaginary value $\frac{i\hbar^2}{2m}$. They had found that
there exists an essential coincidence between two presentations (\ref{b1}) and
(\ref{b2}) if the coefficient of the diffusion $D$ is equal of $\frac{i\hbar}
{2m}$. Therefore Feynman has used for transition between two OrbWvFncs $\Psi $
of some free FrthQntMcrPrt with different coordinates and times the following
formula:
\begin{equation}
\label{c1}
\Psi (x_1,t_1)\,=\,\int\,K(x_1,t_1|x_2,t_2)\,\Psi(x_2.t_2)\,dx_2
\end{equation}

in analogous of such the formula,which early had been used by Einstein (\cite
{AE}),\cite{ES}),Smoluchovski (\cite{MS}) and Wiener (\cite{NW}) for the
transition between two DstFncsPrbDns $W(\lambda ,t)$ of a free BrnClsMicrPrt:
\begin{equation}\label{c2}
W(\lambda ,t)\,=\,\int\,W({\lambda}_o,t_o)\,P({\lambda}_o,t_o|\lambda ,t)
\,d{\lambda}_o\,
\end{equation}

 The diffusivity $D$, which is very strongly dependent as on the viscosity
and temperature of the solvent, so on the radius of the BrnClsPrt, can been
determined by help of the DstFncsPrbDns $W(\lambda,t)$ by means means of
the followig definition formula :
\begin{equation}
\label{c3}
D(\lambda_o)\,=\,{\L\,i\,m \choose \Delta t \rightarrow 0}\,
\cdots \int_a^b\,\frac{(\lambda\,-\,\lambda_o)^2}{2\Delta t}\,
P(\lambda_o,t_o|\lambda ,t)\,d\lambda
\end{equation}

It is neccessary to turn here our attention to satisfaing as from the
functions of the hit probability $P(x_1,t_1|x_3,t_3)$ and $P(x_3,t_3|x_2,t_2)$, so from the functions $K(x_1,t_1|x_3,t_3)$ and
$K(x_3,t_3|x_2,t_2)$ of the following M-change relations,which characterizes
Markovian processes:
\begin{equation}
\label{d1}
P(x_1,t_1|x_2,t_2)\,=\,\int P(x_1,t_1|x_3,t_3)\,P(x_3,t_3|x_2,t_2)\,dx_3
\end{equation}

and
\begin{equation}
\label{d2}
K(x_1,t_1|x_2,t_2)\,=\,\int K(x_1,t_1|x_3,t_3)\,K(x_3,t_3|x_2,t_2)\,dx_3
\end{equation}

if the probability function $K$ for the FrthQntMicrPrt within the NrlQntMch
has the following well known form :
\begin{equation}
\label{d3}
K(x_o,t_o|x,t)\,=\,\frac{\sqrt{m}}{\sqrt{2i\pi \hbar t}}\,
\exp[-i\,\frac{mx^2}{2\hbar t}]
\end{equation}

which is analogous of the probability function P for the BrnClsMcrPrt within
the StchClsMch having the following Gaussina exponential form :
\begin{equation}
\label{d4}
P(x_o,t_o|x,t)\,=\,\frac{1}{\sqrt{4\pi \tau D}}\,
\exp[-\,\frac{x^2}{4\pi D\tau}]
\end{equation}

 But as we can see from (\ref{d3}) and (\ref{d4}) that one have no physical
mean of some classical velocity, as if
\begin{equation} \label{d5}
{\L\,i\,m \choose t_{n} \rightarrow t_{n-1}}\,\ldots\,
\left(\frac{x_n\,-\,x_{n-1}}{t_n\,-\,t_{n-1}}\right)\,\ne\,
{\L\,i\,m \choose t_{n+1} \rightarrow t_{n}}\,\ldots\,
\left(\frac{x_{n+1}\,-\,x_{n}}{t_{n+1}\,-\,t_{n}}\right)\,
\end{equation}

when the BrnClsMicrPrt participates within the BrnStchMtn. From here it
follows that although Feynman speak very loudly about his using of the
smallest action principle at description of the unknown uncommon behaviour of
the QntMicrPrts, in a reality he go on very silently by dint of the
matematical apparatus of the BrnStchMch, using the existent substantial
analogy between the FrthStchMtn of the QntMicrPrt and well known BrnStchMtn
of the BrnMicrPrt.

 As a generalization of the equalityes (\ref{d1}) and (\ref{d2}) they have
shown that the probability function describes the probabiility of some free
QntMacrPrt (BrnClsPrt) to move from the point $x_o$ in the time moment $t_o$
to the point x in the time t, passing through the interval of some virtual
trajectory between the points a and b ,is clearly defined as a product from
the probaibility of same free BrnClsPrt to move from the point $x_o$ in the
time moment $t_o$ to the point $x_1$ in the time moment $t_1$, passing through
the interval of some virtual trajectory between the points $a_1$ and $b_1$,
times the probability of same free BrnClsPrt to move from the point $x_1$ in
the time moment $t_1$ to the point $x_2$ in the time moment $t_2$, passing
through the interval of some virtual trajectory between the points $a_1$ and
$b_1$, times the probability of same free BrnClsPrt to move from the point
$x_2$ in the time moment $t_2$,passing through the interval of some virtual
trajectory between the points $a_2$ and $b_2$ and so on,times the probability
of same free BrnClsPrt to move from the point $x_n$ in the time moment $t_n$,
passing through the interval of some virtual trajectory between the points
$a_n$ and $b_n$, after their integration in respect of all the intermediate
variables over their intervals :
\begin{eqnarray} \label{d6}
P(x_o,t_o|x,t)\,=\,\int\ldots\int\,P(x_o,t_o|x_1,t_1)\,P(x_1,t_1|x_2,t_2)
\nonumber\\
P(x_2,t_2|x_3,t_3)\,\ldots\,P(x_n,t_n|x,t)\,dx_1,dx_2,dx_3,\ldots\,dx_n
\end{eqnarray}

and analogous
\begin{eqnarray}
\label{d7}
K(x_o,t_o|x,t)\,=\,\int\ldots\int\,\,K(x_o,t_o|x_1,x_1)\,K(x_1,t_1|x_2,t_2)
\nonumber \\
K(x_2,t_2|x_3,t_3)\,\ldots\,K(x_n,t_n|x,t)\,dx_1,dx_2,dx_3,\ldots\,dx_n
\end{eqnarray}

 However,it is very important to understand why the form of probability
function $K_{i,j}$ of two independent events,having property of a product of
their own probability function $K_i$ and $K_j$, has exponential connection
with the action function $S_{i,j}(r,t)$ of a free QntMicrPrt, having property
of a sum of two independent events. Therefore the form (\ref{d3}), written by
Feynman, coinsidences with the gaussian exponent (\ref{d4}), very early writing
down for description of the probability $P(r_o,t_o|r,t)$ to find some BrnClsPrt
after a time interval $\tau\,=\,t\,-\,t_o$ of a distance $x\,=\,r\,-\,r_o$. Hence
if the DstFncPrbDns $W$ may has positive real value only, the OrbWvFnc $\Psi $
may has a complex value. This means that for some part of the OrbWvFnc $\Psi $
may exist a total analogy between both the QdrPrtEqn and their solutions. Indeed,
if the exponent and normalization factor of the DstFncPrbDns $W$ have real value
only, the exponent and normalization factor of the OrbWvFnc $\Psi $ may have
complex value.

 Indeed,if we suppose that
\begin{equation}
\label{d8}
 E\,=\,\frac{\{\bar p\}^2}{2m}\,+\,\frac{\{\delta p\}^2}{2m}\,+\,U(r)\,
\end{equation}

then within a quasiclassical approximatin many physicists presume that the
SchrEl's OrbWvFnc $\Psi $ may been written in the following two forms :
a) within the classical accessible area :
\begin{equation}
\label{d9}
\Psi(r,t)\,=\,\left\{\,C_1\,\exp\{\frac{i}{\hbar}\,\int\,p\,dx\}\,
\exp\{-\frac{1}{2}\ln p\}\,+
\,C_2,\exp\{\frac{-i}{\hbar}\,\int\,p\,dx\}\,
\exp\{-\frac{1}{2}\ln p\}\,\right\}
\end{equation}

and b) within the classical accessible area :
\begin{equation}
\label{d10}
\Psi(r,t)\,=\,C_1\,\exp\{-\frac{1}{\hbar}\,\int\,p\,dx\}\,\exp\{-\frac{1}{2}
\ln p\}
\end{equation}

where
\begin{equation}
\label{d11}
p\,=\,\sqrt{2m\{E\,-\,U\}}\,=\,\sqrt{\{\bar p\}^2\,+\,\{\delta p\}^2}\,
\end{equation}

if $\bar p\,=\langle\,|p|\,\rangle\,$ \quad and \quad $\{\delta p\}^2\,=
\,\{p\,-\,\bar p\}^2 $.  When $ U\,\ge\,E$ then
\begin{equation}
\label{d12}
|p|\,=\,i\sqrt{2m\{U\,-\,E\}}\,=\,i\,\sqrt{\{\bar p\}^2\,+ \,\{\delta p\}^2}
\end{equation}

 Then the imaginary part of the exponent (the real part $S_1$ of the action
function $S$),
\begin{equation}
\label{d13}
\{\frac{i}{\hbar}\}\,\int\,p\,dx\}\,
\quad and \quad
\{\frac{-i}{\hbar}\}\,\int\,p\,dx\}
\end{equation}

will describe the classical motion along distinct smooth narrow line and the
real part of the exponent (the imaginary part $S_2$ of the action function $S$)
\begin{equation}
\label{d14}
\frac{1}{2}\ln(p)
\end{equation}

will describe the Furthian stochastic motion (FrthStchMtn) along a frequently
broken of petty strongly disorientated small pieces closely to the smooth and
distinc thin line of the classical motion.  Therefore the FrthStchMtn will
erode the clear-cut smooth thin line and the total motion of the QntMcrPrt
will be spread in a wide path. As the energy E, the averaged momentum $\bar p
$ and the dispersion $\sqrt{(\delta p)^2}$ are determined by the OrbWvFnc
$\Psi $  of the QntMicrPrt then the possibility decrease of its discovery
within the appointed area will be also determined. When the potential $U$ is
bigger then the total energy $E$ of some QntMicrPrt, then the momentum $p$
must be substituted by the $\,i\,|p|$. As a result of that we can suppose
that the unusual dualistic behaviour of the QntMicrPrt within the NrlQntMch
can be described by dint of the following mutual conjugated physical
quantities :
\begin{equation}
\label{e1}
r_j\,=\,\bar {r_j}\,+\,\delta {r_j}\quad and\quad
p_j\,=\,\bar {p_j}\,+\,\delta {p_j}
\end{equation}

 The upper supposition shows us way for some part of the QntMicrPrt's OrbWvFnc
may exist a total analogy between the presentations of both the QdrPrtDfrEdts
and their solutions.In this way we understand why the behaviour of the
QntMicrPrt must be described by a OrbWvFnc $\Psi $ , although the behaviour
of the ClsMacrPrt may be described only by a clear-cut smooth thin line.

Indeed, the first, it is known from quantum electrodynamics (QntElcDnm), that
when the energy of some QntMicrPrt has a complex value, then its real part
describes the real energy of the particle, while its imaginary part describes
its disappearance in the time, i.e.the time of its decay into another
QntMicrPrts; in the second, it is known from quantum mechanics theory
(QntMchThr) of the Solid State, that the real part of the momentum of the
QntMicrPrt describes its averaged current part, while the imaginary part of
the momentum of the QntMcrPrt describes its disappearance in the space, i.e.
the decrement of the probability the QntMcrPrt to come in inside of the
potential barier $U$.

 Hence, although that Feynman speak loudly about the principle of the smallest
action function, but he uses always the mathematical apparatus of the Brownian
stochastic motion (BrnStchMtn), as the imaginary part of the action $S$ of the
FrthQntMcrPrt takes the form of the real part of the exponent of the DstFncPrbDns
$W$ of BrnClsMicrPrt, which describes its BrnStchMtn.We can impressively see this
discrepancy between interpretation and using the mathematical apparatus of the
ClsStchMch, particulary at the derivation of the Schrodinger wave equation by
using of some formulas from the BrnStchMtn theory with the consideration the
potential role.
\begin{equation}
\label{e2}
\Psi (x,t\,+\,\epsilon)\,=\,\frac{1}{A}\,\int_V\,\exp[\frac{im\eta^2}
{2\hbar\epsilon}]\,\Psi ({x\,+\,\eta},t)\,d\eta
\end{equation}

Indeed, if Feynman has used in a reality the principle of the smallest action,
he would not have expand in a power only the potential exponent and keeping of
this part of the action, which describes the kinetic energy of the SchEl's
FrthStchMtn. But Feynman whould expand the Lagrangian exponent, as there is
distinction between the kinetic and potential energies, which in according
with the smallest action principle must compensate each other. The expansion
of the potential exponent only :
\begin{equation}
\label{e3}
\exp[\frac{i\epsilon}{\hbar}\,U(x\,+\,\frac{\eta}{2},t\,)]\,\cong\,[\,1\,-
\,\frac{i\epsilon}{\hbar}\,U(x\,+\,\frac{\eta}{2},t\,)]
\end{equation}

means that Feynmam keeps the kinetic energy exponent $exp[\frac{im\eta^2}
{2\hbar\epsilon}]$ for averaging the interaction of the FrthQntMcrPrt by means
of the DstFncPrbDns $W$, assuming that the QntMicrPrt perform the
FrthQntStchMtn. I think that Feynman has expanded the potential exponent
because it compensates the kinetic energy of the NtnClsMtn, which participates
in a QvdDfrClsEqtHml-Jcb and as last it would break semi-group properties of
the Gausian exponential distribution.

\section{\normalsize Calculation of the minimal dispersions of some dynamical
variables of QntMicrPrts as a result of their participation in the
FrthQntStchMtn}

We attempt in what follows to show that the smalles values of some dynamical
variable dispersions may been determined as a result of their participation
in the FrthQnttchMtn,using their definition by Feynman.Hence when Feynman has
discussed about the time dependence of the velocity of some QntMicrPrt
\begin{equation}
\label{C}
v_n^+\,=\,{\L\,i\,m \choose \epsilon\,\to\,0}\,
\frac{\left(x_{n+1}\,-\,x_n\right)}{\left(t_{n+1}\,-\,t_n\right)}\,=\,(\,v\,+\,iu) ,
\end{equation}

and
\begin{equation}
\label{D}
v_n^-\,=\,{\L\,i\,m \choose \epsilon\,\to\,0}\,
\frac{\left(x_n\,-\,x_{n-1}\right)}{\left(t_n\,-\,t_{n-1}\right)}\,=\,(v\,-\,iu) ,
\end{equation}

where
\begin{equation}
\label{F}
t_{n+1}\,=\,t_n\,+\,\epsilon \quad and \quad t_{n-1}\,=\,t_n\,-\,\epsilon,
\end{equation}

, he has assumed that it is
\begin{equation}
\label{G}
u^2\,\approx\,\left(\frac{2D}{\epsilon}\right)
\end{equation}

which may be only if
\begin{equation}
\label{H}
\mid\,x_{n+1}\,-\,x_{n}\,\mid\,\cong\,\mid\,x_{n}\,-\,x_{n-1}\,\mid\,\approx\,
\sqrt{\left(2D\epsilon\right)}\,=\,\sqrt{\left(\frac{\hbar\epsilon}{m}\right)},
\end{equation}

, as it must be at FrthStchMtn. Indeed, if
\begin{equation}
\label{K}
\langle(\Delta p)^2\rangle)=\frac{1}{2}(\Delta p)^2\,=\,\frac{(m\Delta x)^2}
{2(\epsilon)^2}\,=\,\frac{m\hbar}{2\epsilon},
\end{equation}

, and if
\begin{equation}
\label{L}
\langle(\Delta x)^2\rangle\,=\,\frac{1}{2}(\Delta x)^2 ,
\end{equation}

, then
\begin{equation}
\label{M}
\langle(\Delta p)^2\rangle\times\langle(\Delta x)^2\rangle\,=\,
\left(\frac{\hbar}{2}\right)^2 ,
\end{equation}

 Further when Feynman has discussed about the kinetic energy of the QntMcrPrt,
he has asserted that instead the known expression:
\begin{equation}
\label{M}
\frac{m(x_{n+1}-x_n)^2}{(2\epsilon)^2}+ \frac{m(x_n-x_{n-1})^2}{(2\epsilon)^2}
\end{equation}

we must use the following expression:
\begin{equation}
\label{P}
2\frac{m}{2}\times\frac{(x_{n+1}\,-\,x_n)}{\epsilon}\times
\frac{(x_n\,-\,x_{n-1})}{\epsilon} ,
\end{equation}

Indeed, if from eqns.(\ref{C}) and (\ref{D}) we have : $v^+ =(v+iu)\quad$ and
$\quad v^- =(v-iu) $ then
\begin{equation}
\label{Q}
\left(\frac{m(x_{n+1}\,-\,x_n)^2}{(2\epsilon)^2}\right)\,+\,
\left(\frac{m(x_n\,-\,x_{n-1})^2}{(2\epsilon)^2}\right)\,=\,
\frac{mv^2}{2}\,-\,\frac{mu^2}{2},
\end{equation}

which is wrong, but
\begin{equation}
\label{R}
\frac{m}{2}\times\frac{(x_{n+1}\,-\,x_n)}{\epsilon}\times
\frac{(x_n\,-\,x_{n-1})}{\epsilon}\,=\,\frac{mv^2}{2}+\frac{mu^2}{2} ,
\end{equation}

, which is correct. Moreover, if both
\begin{equation}
\label{S}
(\Delta E)\,=\,\frac{mu^2}{2}
\end{equation}

, where
\begin{equation}
\label{T}
u^2\,=\,\frac{\langle (\Delta x)^2\rangle )}{(\epsilon )^2}\,=\,
\frac{(\Delta x)^2}{2(\epsilon)^2}\,=\,\frac \hbar {2m\epsilon},
\end{equation}

and from (\ref{F}) $\quad \Delta t\,=\,\epsilon $ then we have immediately :
\begin{equation}
\label{V}
\langle(\Delta E)^2\rangle\times\langle(\Delta t)^2\rangle\,\,=\,
\left(\frac{\hbar}{2}\right)^2 .
\end{equation}

 Further the values of the dispersion $\langle(\Delta {P_r})^2\rangle\,$ and
$\langle(\Delta {L_j})^2\rangle\,$ can been determined by virtue of the
uncertainity relations of Heisenberg :
\begin{equation}
\label{W1}
\langle(\Delta {P_r})^2\rangle\,\times\,
\langle(\Delta r)^2\rangle\,\ge\,\frac{\hbar^2}{4}
\end{equation}

\begin{equation}
\label{W2}
\langle(\Delta {L_x})^2\rangle\,\times\,\langle(\Delta {L_y})^2\rangle\,\ge\,
\frac{\hbar^2}{4}\times\,\langle(L_z)^2\rangle
\end{equation}

\begin{equation}
\label{W3}
\langle(\Delta {L_y})^2\rangle\,\times\,\langle(\Delta {L_z})^2\rangle\,\ge\,
\frac{\hbar^2}{4}\times\,\langle(\Delta {L_x})^2\rangle
\end{equation}

and
\begin{equation}
\label{W4}
\langle(\Delta {L_z})^2\rangle\,\times\,\langle(\Delta {L_x})^2\rangle\,\ge\,
\frac{\hbar^2}{4}\times\,\langle(\Delta {L_y})^2\rangle
\end{equation}

 Thence the distersion $\langle(\Delta {P_r})^2\rangle$ will really have its
minimal value at the maximal value of the $\langle(\Delta r)^2\rangle$, which
is $\approx\,\langle\,r^2\,\rangle$. In such a way we obtained that the
minimal value of the dispersion $\langle(\Delta {P_r})^2\rangle$ can been
determined by the following well known inequality :
\begin{equation}
\label{X1}
\langle(\Delta {P_r})^2\rangle\,\ge\,
\frac{\hbar^2}{4}\,\times\,\langle(\Delta r)^2\rangle\,
\end{equation}

 When the SchEl is placed within an external potential with the cylyndrical
symmetry then the direction of the axis z of the our coordinate system
coincideswith the direction of the angular MchMmn,then $\langle(L_z)\rangle\,$
therefore by means of (\ref{W2}) we can obtain that :
\begin{equation}
\label{X2}
\langle(\Delta {L_x})^2\rangle\,=\,\langle(\Delta {L_y})^2\rangle\,=\,
\frac{l\hbar^2}{2}
\end{equation}

 As by means of the inequalities (\ref{W3}) and (\ref{X2}) we can obtain that
$\langle(\Delta {L_z})^2\rangle\,\approx\,\frac{\hbar^2}{4}$, then we can
obtain that: at $\langle(L_z)\rangle\,\ne\,0$,i.e. at an existence of the
cylindrical symmetry $(\langle(\Delta {L_x})^2\rangle\,=\,
\langle(\Delta {L_y})^2\rangle\,=\,\frac{l\hbar^2}{2}$
\begin{equation}
\label{X3}
\langle(L)^2\rangle\,=\,(\langle{L_z}\rangle)^2\,+\,\langle(\Delta {L_x})^2
\rangle\,+,\langle(\Delta {L_y})^2\rangle\,+\,\langle(\Delta {L_z})^2\rangle\,
=\,(l\hbar)^2\,+\,l\hbar^2\,+\,\frac{\hbar^2}{4}\,=\,(\,l\hbar\,+\,
\frac{\hbar}{2}\,)^2
\end{equation}

and at $\langle{L_z}\rangle\,=\,0$,i.e. at an existence of the spherical
symmetry $(\langle(\Delta {L_x})^2\rangle\,=\,\langle(\Delta {L_y})^2\rangle\,
=\, \langle(\Delta {L_z})^2\rangle\,=\,\frac{\hbar^2}{4})$
\begin{equation}
\label{X4}
\langle(\Delta {L_x})^2\rangle\,+,\langle(\Delta {L_y})^2\rangle\,+\,
\langle(\Delta {L_z})^2\rangle\,=\,\frac{3\hbar^2}{4}
\end{equation}

 Realy, the obtained upperesults may been obtained by means of the formal
transfer from the three-dimensional QdrPrtDfrWvEqtSchr for the spherical part
$R(r)$ of the SchrEl's OrbWvFnc $\Psi $,depend only from $r$, written in a
spherical coordinate system:
\begin{equation}
\label{Y1}
\frac{d^2 R}{dr^2}\,+\,\frac{2}{r}\,\frac{dR}{dr}\,+\,\left[\frac{2m}{\hbar^2}
\,\left\{\,E\,-\,U(r)\,\right\}\,-\,\frac{l(l+1)}{r^2}\,\right]\,R(r)\,=\,0
\end{equation}

to two dimensional QdrPrtDfrWvEqtSchr for the cylindrical part $\Phi(\rho)$ of
the SchrEl's OrbWvFnc $\Psi $, depend only from $\rho $, written in a
cylindrical coordinate system :
\begin{equation}\label{Y2}
\frac{d^2 \Phi}{d(\rho)^2}\,+\,\frac{1}{\rho}\,\frac{d\Phi}{d\rho}\,+
\,\left[\frac{2m}{\hbar^2}\,\left\{\,E\,-\,U(\rho)\,\right\}\,-
\,\frac{(l+1/2)^2}{(\rho)^2}\,\right]\,\Phi(\rho)\,=\,0
\end{equation}

 There is necessity to point here, that the formal transfer from the equation
(\ref{Y1}) to the equation (\ref{Y2}) can been realized by virtue of the
exchange of $r$ and $R(r)$ with $\rho$ and $\frac{\Phi(\rho)}{\sqrt{\rho}}$ in
the corresponding way. Further it is well known that the presentation of the
QdrPrtDfrWvEqtSchr for the SchEl's total OrbWvFnc $\Psi(\rho,\varphi,z)$ has
the following well known form :
\begin{equation}
\label{Y3}
\frac{\partial^2 \Psi}{\partial \rho^2}\,+
\,\frac{1}{\rho}\,\frac{\partial \Psi}{\partial \rho}\,+
\,\frac{1}{\rho^2}\,\frac{\partial^2\Psi}{\partial \varphi^2}
\,+\,+\,\frac{\partial^2 \Psi}{\partial z^2}\,+\,\frac{2m}{\hbar^2}\,
\left\{\,E\,-\,U(\rho,z)\,\right\}\Psi\,=\,0
\end{equation}

 Then in a result of the comparison of the eq.(\ref{Y2}) with the eq.(\ref{Y3}) we can obtain the averaged value of the total OrbMchMnt's square
$<L^2>$ in the NrlQntMch must coincide with its value $\hbar(l+1/2)$
determined by the eq. (\ref{X3}).In such a way we obtain the average value of
the total orbital mechanical moment (OrbMchMmn) in a square $\langle\,L^2\,
\rangle$ of a SchEl in the cylindrical coordinate.From above it is followed
that the value $\langle\,L^2\,\rangle\,=\,\hbar^2 l(l+1)$, which can be
obtained in the spherical coordinate, taking no into account the part
$\langle\,(\Delta {P_r})^2\rangle\,=\,\frac{\hbar^2}{4r^2}$ .Indeed,after all
the part of the product $\langle(\Delta {P_r}^2)\rangle\,\times\,
\langle\,r^2\,\rangle$ may be considered as the dispersion
$\langle\,(\Delta {L_r})^2\,\rangle$ of $\langle\,(L_r)^2\,\rangle$
along axis $r$, when the axis $z$ coincides with the radius-vector $r$.
In such a way we can write the SchEl's OrbWvFnc as a result of an upper
discussion in the following presentations :
\begin{equation}
\label{Y4}
\Psi (\rho,\varphi,z)\,=\,\Psi_l(\rho,\varphi,z)\,\exp(i\frac{\varphi}{2})\,=
\,\Psi_l(\rho,z)\exp(i\frac{(2l\,+\,1)\varphi}{2})
\end{equation}

\section{\normalsize Conclution}

 The realized above investigation shows that when the SchEl is moving in
Coulomb potentiale of the NclElcChrg of some H-like atom, then the stability
of its ground state is ensured by the existence of the SchEl's kinetic energy
of its FrthStchMtn, generated as a result of the continuous ElcIntAct  of its
BlrElmElcChrg with QntElcMgnFld of stoch astic created virtual photons
(StchVrtPhtns) within the fluctuating vacuum (FlcVcm). Besides that it could
be easily shown that not only the SchEl's localized energy, ensuring a
stability of its ground state within H-atoms, but as well as all those are
following: the existence of its additional MchMm and MgnDplMm,the SchEl's
tunnelling through the potential barrier and the shifts of its energy level
in an atoms - are natural and incontestable manifestations of its effective
participation in the FrthStchMtn too. However there exist an essential
difference between Brownian classical stochastic motion (BrnClsStchMtn) of
some BrnClsPrt within NrlClsMch and Furth's quantum stochastic motion
(FrthQntStchMtn) of some QntMicrPrt within NrlQntMch as the result of the
existent difference between both moving cause: stochastic scattering of some
atoms or molecules from another BrnClsPrt and the ElcIntAct of the SchEl's
BlrElmElcChrg with the ElcInt of LwEn-StchVrtPhtns, generated stochasticaly
in the FlcVcm through continuous exchanges of the LwEn-StchVrtPhtns between
the SchEl's WllSpr ElmElcChrg and the FlcVcm .

 In such a natural way we had ability to obtain the minimal value of the
dispersion product, determined by the Heisenberg uncertainty relation. Hence
we can come to a conclusion that the dispersions of the dynamical parameters
of the quantized micro particles are natural result of their forced motions
owing to ElcMgnIntAct of its WllSpr ElmElcChrg or MgnDplMm with the resultant
strengths of the ElcFld or MgnFld of QntElcMgnFlds of StchVrtPhtns at its
FrthQntStchMtn through the FlcVcm.

\end{document}